\documentclass[12pt]{article}
\usepackage{graphicx}
\usepackage{epsfig}
\usepackage{psfrag}
\usepackage{latexsym}
\usepackage{indentfirst}
\usepackage{fancyhdr}
\usepackage{amsmath,amssymb,amsfonts}
\usepackage{amscd}
\usepackage{color}

\newcommand{\pa}{\partial}

\newcommand{\be}{\begin{equation}}
\newcommand{\ee}{\end{equation}}
\newcommand{\bc}{\begin{center}}
\newcommand{\ec}{\end{center}}
\newcommand{\bea}{\begin{eqnarray}}
\newcommand{\eea}{\end{eqnarray}}

\newcommand{\nn}{\nonumber}
\newcommand{\la}{\label}

\begin{document}
\title{\bf Schwarzschild-de Sitter Metric and Inertial Beltrami Coordinates}

\author{Li-Feng Sun, Mu-Lin Yan\footnote{E-mail address: mlyan@ustc.edu.cn}, Ya Deng,
Wei Huang\footnote{Corresponding author. E-mail address: weihuang@mail.ustc.edu.cn},
Sen Hu\footnote{E-mail address: shu@ustc.edu.cn}\\
Wu Wen-Tsun Key Lab of Mathematics of Chinese Academy of Sciences,\\
School of Mathematical Sciences, and Department of Modern Physics,\\
University of Science and Technology of China, Hefei, Anhui 230026, China}
\maketitle

\begin{abstract}
Under consideration of coordinate conditions, we get the Schwarzschild-Beltrami-de Sitter (S-BdS) metric solution of the Einstein field equations with a cosmological constant $\Lambda$. A brief review to the de Sitter invariant special relativity (dS-SR), and de Sitter general relativity (dS-GR, or GR with a $\Lambda$) is presented. The Beltrami metric $B_{\mu\nu}$ provides inertial reference frame for the dS-spacetime. By examining the Schwarzschild-de Sitter (S-dS) metric $g_{\mu\nu}^{(M)}$ existed in literatures since 1918, we find that the existed S-dS metric $g_{\mu\nu}^{(M)}$ describes some mixing effects of gravity and inertial-force, instead of a pure gravity effect arisen from ``solar mass'' $M$ in dS-GR. In this paper, we solve the vacuum Einstein equation of dS-GR, with the requirement of gravity-free metric $g_{\mu\nu}^{(M)}|_{M\rightarrow 0}=B_{\mu\nu}$. In this way we find S-BdS solution of dS-GR, written in inertial Beltrami coordinates. This is a new form of S-dS metric. Its physical meaning and possible applications are discussed.

\vskip0.1in
\noindent PACS numbers: 04.20.Jb; 11.30.Cp; 98.80.Jk

\noindent Key words: Classical general relativity, Exact solutions, Special Relativity, de Sitter spacetime symmetry, Beltrami metric, Mathematical and relativistic aspects of cosmology.
\end{abstract}

\newpage

\section{Introduction}\label{sec:intr}

\noindent
Discussions to de Sitter(dS) spacetimes have attracted much interests recently. The reasons are multiple. Two of them are: (1) The recent
observations in cosmology show that our universe is in accelerated expansion (see, e.g., \cite{peebles} and references within), which implies that the universe is probably asymptotically dS spacetime with positive cosmological
constant $\Lambda$; (2) Just as weakening the fifth axiom leads to non-Euclidean
geometry, giving up Einstein's Euclidean assumption on the rest
rigid ruler and clock in special relativity leads to other kind
of Special Relativity (SR) on the dS-spacetime with dS-radius $R$
\cite{look,Lu74,Guo1,Guo2,Guo3,Guo4,Ours,Yan1,Yan2,Yan3}. We call it dS-SR. Localizing the spacetime symmetry in inertial frames of dS-SR, we can reach gravitational field theory with a cosmological constant $\Lambda\equiv 3/R^2$ \cite{Yan2,Yan3}. Such theory is just de Sitter General Relativity (dS-GR) with $\Lambda$. In this paper we try to solve the equation of dS-GR in vacuum,
and discuss Schwarzschild-de Sitter metric in inertial
Beltrami spacetime coordinates.\footnote{See Appendix,
the Beltrami coordinate system is defined by the ratio of two homogeneous coordinates
(say, $x^\mu = R\xi^\mu/\xi^5$) when the de Sitter spacetime,
as a hypersurface, is embedded in a homogeneous spacetime
spanned by $\xi^A$, $A=\{0,1,2,3,5\}$.}

Existence of inertial coordinate system is the foundation of special relativity.
And existence of local inertial coordinate system is one of GR-principles.
What we say the inertial coordinate system here is that in which the inertial motion law for free particles holds. Namely, in a maximally symmetric spacetime with a specific metric $g_{\mu\nu}$, if the free particle motion is inertial, we could call such sort of $g_{\mu\nu}$ {\it inertial metric}.
There are two inertial metrics: Minkowski spacetime metric $\eta_{\mu\nu}=diag\{1,-1,-1,-1\}$, and Beltrami metric (see Appendix Eq.(\ref{3-4})). It is easy to check $\eta_{\mu\nu}$ is inertial. The Landau-Lifshitz action \cite{Landau} for free particle with mass $m_0$ in Minkowski spacetime is:
\begin{equation}\label{y01}
A=-m_0c^2\int ds=-m_0c^2\int \sqrt{\eta_{\mu\nu}dx^\mu dx^\nu}:=\int dt L,
\end{equation}
and then
\begin{equation}\label{y02}
L=-m_0c^2\sqrt{1-\frac{\dot{\mathbf{x}}^2}{c^2}}\equiv L(\dot{\mathbf{x}}).
\end{equation}
From $\delta A=0$ (the least action principle, or the free-particle motion law along geodesic line), the equation of motion of the particle reads
\begin{equation}\label{y03}
\frac{d}{dt}\frac{\partial L(\dot{\mathbf{x}})}{\partial\dot{\mathbf{x}}}=\frac{\partial L(\dot{\mathbf{x}})}{\partial \mathbf{x}}~\Rightarrow~ \ddot{\mathbf{x}}=0,
\end{equation}
and then $\dot{\mathbf{x}}=constant$. We conclude that the the inertial motion law for free particle holds in Minkowski spacetime, and metric $\eta_{\mu\nu}$ is inertial. Similarly, it can also be proved that the Betrami metric $B_{\mu\nu}(x)$ is also inertial via straightforward calculations (see Appendix: Eqs.(\ref{A6})-(\ref{B10})). The fact that $B_{\mu\nu}(x)$ is inertial led to the discovery of de Sitter invariant special relativity \cite{look,Lu74,Ours}.

Physically, it is useful and meaningful to find out GR-solutions for empty spacetime, which approach to the metric of the inertial system when the gravity vanishes. A typical example is usual Schwarzschild solution of GR without $\Lambda$. To empty spacetime with $T_{\mu\nu}=0$, the Einstein equation reads
\begin{eqnarray}\label{1}
\mathcal{R}_{\mu\nu}=0,
\end{eqnarray}
where $\mathcal{R}_{\mu\nu}$ is Ricci tensor. The Schwarzschild solution of (\ref{1}) in spherical space coordinates $r,\;\theta,\;\phi$ is
\begin{eqnarray}\label{2}
ds^2=\left(1-\frac{2GM}{c^2r}\right)c^2dt^2
-\left(1-\frac{2GM}{c^2r}\right)^{-1}dr^2-r^2(d\theta^2+\sin^2\theta d\phi^2),
\end{eqnarray}
where $M$ is ``solar" mass. It is essential that when $M\rightarrow 0$, the Schwarzschild metric approaches Minkowski ($Mink$) metric, which provides inertial reference frames. Explicitly, when $M\rightarrow 0$, from (\ref{1}), we have
\begin{eqnarray}\label{3}
ds^2\rightarrow ds_{Mink}^2=c^2dt^2-dr^2-r^2(d\theta^2+\sin^2\theta d\phi^2).
\end{eqnarray}
(Note, in Cartesian space coordinates $\{x^1,\;x^2,\;x^3\}$,
$ds_{Mink}^2=\eta_{\mu\nu}dx^\mu dx^\nu$ where
$\{\eta_{\mu\nu}\}=diag\{1,-1,-1,-1\}$ and $x^0=ct$. $\eta_{\mu\nu}$ is a solution of (\ref{1})). This fact indicates that when the gravity disappears, the spacetime becomes Minkowski's.
Thanks to this outstanding property, one can use the Schwarzschild metric to achieve the calculations of effects such as the motion in a centrally symmetric gravitational field to verify GR (see, e.g.,\cite{Landau}, {\it pp.306}).

From the above we learned that the Schwarzschild solution structures in GR rely on two essential properties:
(a) the metric satisfies the Einstein equation in empty spacetime with $T_{\mu\nu}=0$;
(b) when the gravity disappear due to $M\rightarrow 0$,
the metric tends to empty spacetime-metric of inertial coordinate system.
To dS-GR (or GR with a cosmologic constant $\Lambda$, see Eq.(\ref{10}) in below), differing from (\ref{1}), the corresponding Einstein field equations for empty spacetime are
\bea\la{001}
\mathcal{R}_{\mu\nu}=\Lambda g_{\mu\nu},
\eea
which will be derived below (see Eq.(\ref{14})). Obviously $\eta_{\mu\nu}$ is no longer the solution of (\ref{001}). We should find a metric which satisfies (\ref{001}), and meanwhile the motion of free particle in the spacetime with this metric is inertial.
In \cite{Yan3}, we have obtained the solution of this problem: (see also Appendix A)
\bea\la{002}
g_{\mu\nu}(x)=B_{\mu\nu}(x)={\eta_{\mu\nu} \over \sigma (x)}+{\eta_{\mu\lambda}\eta_{\nu\rho} x^\lambda
x^\rho\over R^2
\sigma(x)^2} ,
\eea
where $R^2=3/\Lambda$ and $\sigma(x)\equiv 1-\eta_{\mu\nu}x^\mu x^\nu/R^2$.
$B_{\mu\nu}(x)$ is called Beltrami metric.
We also call both $\eta_{\mu\nu}$ and $B_{\mu\nu}(x)$ {\it inertial metrics}.
The SR based on $\eta_{\mu\nu}$ is usual Einstein SR (E-SR),
and the one based on $B_{\mu\nu}(x)$ is dS-SR. In this case,
one may ask what is the Schwarzschild-de Sitter metric of dS-GR
written in inertial Beltrami coordinates?
Namely, a metric satisfies both (\ref{001}) and the requirement that when the gravity disappears (corresponding to ``solar mass'' $M \rightarrow 0$) it tends to Beltrami metric $B_{\mu\nu}(x)$.
In general relativity, a suitable choice of the coordinate system
is often useful to solve actual problems or make actual predictions.
Similarly, the metric within asymptotic inertial Beltrami-spacetime frame is a new and useful metric.
We call such metric{Schwarzschild-de Sitter metric in inertial Beltrami coordinates,
or Schwarzschild-Beltrami-de Sitter metric.
The aim of this paper is to solve this problem.

The paper is organized as follows. In section \ref{sec:2}, we briefly review the dS-SR, and construct dS-GR via localizing the global dS spacetime symmetry in dS-SR. We show that Beltrami metric plays an essential role for charactering the inertial systems of dS spacetime;
In section \ref{sec:3}, we reexamine the old Schwarzschild-de Sitter metric existed in literatures since 1918, and show that it contains both effects of gravity and effects of non-inertial forces. After that we solve the vacuum Einstein equation of dS-GR under the requirement that the metric must purely reflect gravity effect. In other words, our new solution is Schwarzschild-de Sitter metric in inertial Beltrami coordinates.
In section \ref{sec:con}, we sum up the main point of this paper and briefly discuss the physical meaning of our new solution presented in the paper. In Appendix A, more interpretations on Beltrami metrics and dS-SR are presented.

\section{de Sitter Special Relativity, de Sitter General Relativity and Beltrami Metric}\label{sec:2}

We start with a brief review to de Sitter Special Relativity (dS-SR) and de Sitter General Relativity (dS-GR). The Lagrangian for a free particle in dS-SR has been shown in \cite{Ours}: (see Appendix A)
\begin{equation}\label{4}
 L_{dS}=-m_0c \frac{ds}{dt}
 =-m_0c{\sqrt{B_{\mu\nu}(x)dx^\mu dx^\nu}\over dt}=-m_0c{\sqrt{B_{\mu\nu}(x)\dot{x}^\mu \dot{x}^\nu}},
 \end{equation}
where $\dot{x}^\mu=\frac{d}{dt}x^\mu$, $B_{\mu\nu}(x)$ is Beltrami
metric: (see Appendix A)
\begin{eqnarray}\label{5}
B_{\mu\nu}(x)={\eta_{\mu\nu} \over \sigma (x)}+{\eta_{\mu\lambda}\eta_{\nu\rho} x^\lambda
x^\rho\over R^2
\sigma(x)^2} ,~~~{\rm{with}}~~~\sigma(x)\equiv 1-{1\over R^2}
\eta_{\mu\nu}x^\mu x^\nu,
\end{eqnarray}
with constant $R$ the radius of the pseudo-sphere in {\it
dS}-space which is related to the cosmological constant via
$R=\sqrt{3/\Lambda}$. The Euler-Lagrangian equation reads
\begin{eqnarray}\label{6}
   {\pa L_{dS} \over \pa x^i}={d \over dt} {\pa L_{dS} \over \pa \dot{x}^i}.
\end{eqnarray}
Substituting (\ref{4}) into the Euler-Lagrangian equation (\ref{6})
and after a long but straightforward calculation, we obtain \cite{Ours} (see Appendix A)
\begin{equation}\label{7}
\ddot{x}^j=0, ~~ \dot{x}^j={\rm constant}\
\end{equation}
This result indicates that the free particle in the Beltrami
space-time $\mathfrak{B} \equiv \{x^\mu,\;g_{\mu\nu}(x)=B_{\mu\nu}(x)\}$
moves along straight line
and with constant coordinate velocities. Namely the inertial motion
law for free particles holds true in the space-time $\mathfrak{B}$,
and hence the inertial reference frame can be set in $\mathfrak{B}$ (see Appendix A).
The coordinates, which would be used for both dS-SR and dS-GR,
are the inertial Beltrami coordinates $\{x^\mu\}$.

When we transform from one initial Beltrami frame $ x^{\mu}$ to
another Beltrami frame $ \tilde{x}^{\mu}$ with the origin
of the new frame $a^{\mu}$ in the original frame, {the}
transformations between them with 10 parameters are as follows
\begin{eqnarray}\label{8}
x^{\mu} \longrightarrow \tilde{x}^{\mu} &=& \pm \sigma(a)^{1/2} \sigma(a,x)^{-1}
(x^{\nu}-a^{\nu})D_{\nu}^{\mu}, \\
\nonumber D_{\nu}^{\mu} &=& L_{\nu}^{\mu}+R^{-2} \eta_{\nu
\rho}a^{\rho} a^{\lambda} (\sigma
(a) +\sigma^{1/2}(a))^{-1} L_{\lambda}^{\mu} ,\\
\nonumber L : &=& (L_{\nu}^{\mu})\in SO(1,3), \\
\nonumber \sigma(x)&=& 1-{1 \over R^2}{\eta_{\mu \nu}x^{\mu} x^{\nu}},
~~\sigma(a,x)= 1-{1 \over R^2}{\eta_{\mu \nu}a^{\mu}
x^{\nu}}.
\end{eqnarray}
Under this transformation, the metric $B_{\mu\nu}$ is preserved \cite{Ours}:
\begin{equation} \label{01}
 B_{\mu\nu}(x)\longrightarrow~\widetilde{B}_{\mu\nu}(\widetilde{x})={\pa x^\lambda \over \pa
 \widetilde{x}^\mu}{\pa x^\rho \over \pa
 \widetilde{x}^\nu}B_{\lambda\rho}(x)=B_{\mu\nu}(\widetilde{x}).
\end{equation}
The ten parameters in (\ref{8}) are 4 space-time transition parameters $a^\mu$, 3 boost parameters $\beta^i$ and 3 space rotation parameters $\alpha^i$ (Euler angles). They are constants and space-time independent. Therefore the dS-SR transformations (\ref{8}) are global. According to the gauge principle, the localization of global symmetry will yield gauge field theory. As is well known that the external spacetime gauge theory is gravitational field theory \cite{Utiyama,Kibble}. Like to localize the global Poincar\'e (or inhomogeneous Lorentz) group transformation, the global transformation of (\ref{8}) can also been localized via $a^\mu\rightarrow a^\mu(x),\;\beta^i\rightarrow \beta^i(x),\;\alpha^i\rightarrow \alpha^i(x)$. Thus, localized transformation of (\ref{8}) reads
\begin{eqnarray}\label{9}
x^\mu\rightarrow f^\mu(x),
\end{eqnarray}
where $f^\mu(x)$ are four arbitrary functions of $x$. Hence, (\ref{9}) represents a general spacetime coordinates transformation, or curvilinear coordinates transformation. Assuming the spacetime is torsion-free just like Einstein did in GR, the affine connection here is also Christoffel symbol:
$\Gamma^\lambda_{\mu\nu}= \Gamma^\lambda_{\nu\mu}$.

Now let us determine the action of gravity fields $S_G\equiv \int d^4x \sqrt{-g}\;\mathcal{G}(x)$ in empty spacetime, where $\mathcal{G}(x)$ is a scalar. To determine $\mathcal{G}(x)$ we should also consider the fact that the equation of the gravitational field must contain derivatives of the ``potentials" (i.e., $g_{\mu\nu}(x)$) no higher than the second order (just as is the case for the electromagnetic field). From the Riemann geometry, it is found that only $\mathcal{R}$ and trivial constant $\Lambda\equiv constant$ satisfies all requirements. Therefore, $\mathcal{G}(x)=a(\mathcal{R}-2\Lambda)$ where $a$ is also a constant. In {\it Gaussian system of units}, $a=-c^3/(16\pi G)$ where $G=6.67\times 10^{-8} cm^3\cdot gm^{-1}\cdot sec^{-2}$ is the universal gravitational constant. Thus we obtain the action of gauge gravity in empty spacetime:
\begin{eqnarray}\label{10}
S_G=-{c^3\over 16\pi G}\int d^4x\sqrt{-g}(\mathcal{R}-2\Lambda).
\end{eqnarray}
From $\delta S_{G}=0$, we obtain
\begin{eqnarray}\label{11}
\mathcal{R}_{\mu\nu}-{1\over2}g_{\mu\nu}\mathcal{R}+\Lambda g_{\mu\nu}=0.
\end{eqnarray}
dS-SR tells us that to empty spacetime the metric must be Beltrami metric (\ref{5}). Namely, one solution of (\ref{11}) is required to be $g_{\mu\nu}=B_{\mu\nu}$. Then the value of constant $\Lambda$ is determined to be
\begin{eqnarray}\label{12}
\Lambda={3\over R^2},
\end{eqnarray}
and (\ref{11}) becomes
\begin{eqnarray}
\label{13} && \mathcal{R}_{\mu\nu}-{1\over2}g_{\mu\nu}\mathcal{R}+{3\over R^2}g_{\mu\nu}=0,\\
\label{14} {\rm or~}&&\mathcal{R}_{\mu\nu}={3\over R^2}g_{\mu\nu}.
\end{eqnarray}
This is the basic equation of the dS-GR in empty spacetime, which is different from the usual GR's (see (\ref{1})).

\section{Schwarzschild-de Sitter solution of dS-GR in Inertial Beltrami Coordinates}\label{sec:3}
\subsection{Schwarzschild-de Sitter solution in non-inertial system}\label{subsec:31}

The simplest vacuum solution of Einstein's equation with a positive cosmological constant were derived by Kottler (1918), Weyl(1919), Trefftz (1922)\cite{Kottler}. It is actually a spherical
Schwarzchild-de Sitter solution of dS-GR. We call that solution
S-dS metric, which is \cite{Kottler}:
\begin{eqnarray}\label{3-1}
  ds^2&=&g^{(M)}_{\mu\nu}(x_N) dx_N^\mu dx_N^\nu\nonumber\\
  &=&(1-\frac{2GM}{c^2 r_N}-\frac{r_N^2}{R^2})c^2dt^2-(1-\frac{2GM}{c^2 r_N}
  -\frac{r_N^2}{R^2})^{-1}d r_N^2\nonumber\\
  &&-r_N^2(d\theta_N^2+\sin^2\theta_N d\phi_N^2) \label{dSS}
\end{eqnarray}
where $M$ is ``solar" mass, and $\{x^\mu_N\}=\{c t_N,\;r_N,\;\theta_N,\;\phi_N\}$
(subindex N is short for Non-inertial system, which will be proved below)
represent the S-dS spacetime coordinates.
When $M\rightarrow 0$, we get empty de Sitter spacetime metric:
\begin{eqnarray}
  \label{3-17}
  ds^2&=&g^{(0)}_{\mu\nu}(x_N) d x_N^\mu d x_N^\nu\nonumber\\
  &=&(1-\frac{r_N^2}{R^2})c^2d t_N^2-(1-\frac{r_N^2}{R^2})^{-1}dr_N^2
  -r_N^2(d\theta_N^2+\sin^2\theta_N d\phi_N^2)
\end{eqnarray}
Let us explore the question whether the empty de Sitter spacetime metric $g_{\mu\nu}^{(0)}$ is a metric of spacetime with inertial frame or not. Namely, we should pursue whether the motion of free particles in de Sitter spacetime with metric $g_{\mu\nu}^{(0)}$ is inertial or not.

For this, we consider the expression of $g^{(0)}_{\mu\nu}(y)$ in the Cartesian space coordinates $y^i$ with $i=\{1,2,3\}$. From (\ref{3-17}), and noting $y^0=x_N^0=ct_N$,
$y^1=r_N\sin\theta_N \cos\phi_N$, $y^2=r_N\sin\theta_N \sin\phi_N$, $y^3=r_N\cos\theta_N$,
with $\eta_{ij}=diag\{-1,-1,-1\}$, we have
\begin{eqnarray}\label{+1}
ds^2&=&g^{(0)}_{\mu\nu}(y)dy^\mu dy^\nu\\
 &=&\left(1+\frac{\eta_{ij}y^iy^j}{R^2}\right)c^2d t_N^2
+\eta_{ij}dy^idy^j \nonumber\\
&&+\left[\left(1+\frac{\eta_{ij}y^iy^j}{R^2}\right)^{-1}-1\right]{\eta_{lk}\eta_{mn}y^ly^mdy^kdy^n\over \eta_{ij}y^iy^j}.\nonumber
\end{eqnarray}
Note there is no boost in $\{x_N^\mu\rightarrow y^\mu\}$, so it is not a transformation between reference systems. $g^{(0)}_{\mu\nu}(x_N)|_{x_N\rightarrow y}=g^{(0)}_{\mu\nu}(y)$ is
nothing, but only a variable change.
Comparing (\ref{+1}) with (\ref{5}), we find:
\begin{eqnarray}\label{3-19}
g^{(0)}_{\mu\nu}(y)\neq B_{\mu\nu}(y).
\end{eqnarray}
This fact indicates that $g^{(0)}_{\mu\nu}(y)$ is generally not a spacetime
metric of inertial reference systems.
In other words, the coordinates $\{y^\mu\}$ are not an inertial coordinate system.
To be more concrete,
let's see the motion of free particle in S-dS spacetime.
The Landau-Lifshitz Lagrangian $L_N(y^i,\dot{y}^i)$ for a free particle in S-dS is
\begin{eqnarray}\nonumber
&& L_{N}(y^i,\dot{y}^i)=-m_0c \frac{ds}{dt_N}
=-m_0c \frac{\sqrt{g^{(0)}_{\mu\nu}(y)dy^\mu dy^\nu}}{d t_N}\\
\label{+2} &&=-m_0c\sqrt{\left(1-{\mathbf{y}^2\over R^2}\right)-\left[{1\over 1-{\mathbf{y}^2\over R^2}}-1\right]
{(\mathbf{y}\cdot\dot{\mathbf{y}})^2\over c^2\mathbf{y}^2}-{\dot{\mathbf{y}}^2\over c^2}},
 \end{eqnarray}
where $\mathbf{y}=y^1\mathbf{i}+y^2\mathbf{j}+y^3\mathbf{k}$
and $\dot{\mathbf{y}}\equiv d\mathbf{y}/dt_N=\dot{y}^1\mathbf{i}+\dot{y}^2\mathbf{j}+\dot{y}^3\mathbf{k}$.
From $\delta S=-m_0c\;\delta \int ds=\delta\int dt_N L(y^i,\dot{y}^i)=0$, we have
\begin{eqnarray}\label{+3}
   \frac{\pa L_{N}(y^i,\dot{y}^i)}{\pa y^i}=\frac{d}{dt_N}
   \frac{\pa L_{N}(y^i,\dot{y}^i) }{ \pa \dot{y}^i}.
  \end{eqnarray}
Substituting (\ref{+2}) into (\ref{+3}),
we can easily obtain the equation of motion $f(\ddot{y}^i,\dot{y}^i,y^i) =0$. For our purpose, an explicit consideration of one-dimensional motion of the particle is enough. Namely, setting $y^2=y^3=0$ and ignoring the motion of $\mathbf{j}$, $\mathbf{k}$ directions,
the equation of motion $f(\ddot{y}^i,\dot{y}^i,y^i)=0$ becomes:
\begin{equation}\label{non-inertial}
\ddot{y}^1=\frac{y^1}{R^2} \left(c^2\left(1-\frac{(y^1)^2}{R^2}\right)
-\frac{3 (\dot{y}^1)^2}{1-\frac{(y^1)^2}{R^2}}\right)
\neq 0.
\end{equation}
This equation explicitly indicates that there exist inertial forces
in de Sitter spacetime with metric $g_{\mu\nu}^{(0)}$,
which make the particle's acceleration in direction $\mathbf{i}$ to be non-zero,
i.e., $\ddot{y}^1\neq 0$.
Consequently, we conclude that the empty de Sitter spacetime metric $g^{(0)}_{\mu\nu}(x_N)$ is not a metric of spacetime within inertial reference systems.
Or, in short, $g^{(0)}_{\mu\nu}(x_N)$ is non-inertial.

\subsection{Coordinate transformation between non-inertial and inertial systems}\label{subsec:32}

We have shown in the above that the Schwartzschild-de Sitter
metric $g_{\mu\nu}^{(M)}(x_N)$
within asymptotically non-inertial framework described by $g_{\mu\nu}^{(0)}(x_N)$
has been derived by \cite{Kottler} from
%\begin{eqnarray}\label{eq3-2-1}
%\left\{
%  \begin{array}{l}
%    R_{\mu\nu}(x_N)=\frac{3}{R^2} g_{\mu\nu}^{(M)}(x_N),\\
%    g_{\mu\nu}^{(M)}(x_N)\Bigr\lvert_{M\rightarrow 0}=g_{\mu\nu}^{(0)}(x_N).
%  \end{array}
%\right.
%\end{eqnarray}
\begin{eqnarray}\label{eq3-2-1}
  && R_{\mu\nu}(x_N)=\frac{3}{R^2} g_{\mu\nu}^{(M)}(x_N),\\
\label{x2}  && g_{\mu\nu}^{(M)}(x_N)\Bigr|_{M\rightarrow 0}=g_{\mu\nu}^{(0)}(x_N).
\end{eqnarray}
It has been addressed in last subsection that $g_{\mu\nu}^{(0)}(x_N)$ is non-inertial. It is meaningful to find the S-dS metric within asymptotic inertial spacetime frame. Namely, we should solve the equation as follows
\begin{eqnarray}\label{eq3-2-2}
&&  R_{\mu\nu}(x)= \frac{3}{R^2} ~{}^\mathfrak{B}g_{\mu\nu}^{(M)}(x),\\
\label{x3}&&  {}^\mathfrak{B} g_{\mu\nu}^{(M)}(x)\Bigr|_{M\rightarrow 0} =B_{\mu\nu}(x).
\end{eqnarray}
Here, we use $^\mathfrak{B}g_{\mu\nu}^{(M)}(x)$ to denote new S-dS metric
with non-zero ``solar mass'' $M$, which is called Schwarzschild-Beltrami-de Sitter metric hereafter, or S-BdS metric in short. The equation of (\ref{x3}) means that $^\mathfrak{B}g_{\mu\nu}^{(M)}(x)$ must satisfy the requirement that the empty spacetime metric (i.e., the metric $^\mathfrak{B}g_{\mu\nu}^{(0)}(x)$) is inertial. This condition ensures $^\mathfrak{B}g_{\mu\nu}^{(M)}(x)$ to be desired new S-dS metric within
inertial Beltrami coordinates.

Since all of $g_{\mu\nu}^{(M)}(x_N),\;g_{\mu\nu}^{(0)}(x_N)$ and $
^{\mathfrak{B}}g_{\mu\nu}^{(0)}(x)\equiv B_{\mu\nu}(x)$ have already been known,
the reference system transformation $T$ between spacetimes $\{x_N^\mu\}\equiv\{ct_N,\;r_N,\;\theta_N,\;\phi_N\}$ and $\{x^\mu\}\equiv\{ct,\;r,\;\theta,\;\phi\}$ can be derived from the tensor-transformation properties of $g_{\mu\nu}^{(0)}(x_N)$ and $B_{\mu\nu}(x)$ (see (\ref{eq3-2-3}) below). Then the desired result of $^{\mathfrak{B}}g_{\mu\nu}^{(M)}(x)$ will be determined by means of $^{\mathfrak{B}}g^{(M)}=T'g^{(M)}T$.
We sketch the logic in following diagram (\ref{commut}).
\begin{eqnarray}\label{commut}
\begin{CD}
g_{\mu\nu}^{(M)}(x_N)
@> (\ref{x2}) > M\rightarrow 0 >g_{\mu\nu}^{(0)}(x_N)\\
@V T V (\ref{49}) V
@V T V (\ref{eq3-2-3}) V\\
{}^\mathfrak{B} g_{\mu\nu}^{(M)}(x)
@> (\ref{x3}) > M\rightarrow 0 > B_{\mu\nu}(x)
\end{CD}
\end{eqnarray}

Now let us derive the reference system transformation $T$ between spacetimes $\{x_N^\mu\}$ and $\{x^\mu\}$.
In spherical coordinates, and from equations (\ref{3-17}) and (\ref{5}) we have
\begin{eqnarray}
\label{dSspherical}
  g_{\mu\nu}^{(0)}(x_N)&=&
  \left(
    \begin{array}{cccc}
      1-\frac{r_N^2}{R^2} & 0 & 0 & 0\\
      0 & -\frac{1}{1-\frac{r_N^2}{R^2}} & 0 & 0\\
      0 & 0 & -r_N^2 & 0\\
      0 & 0 & 0 & -r_N^2\sin^2\theta_N
    \end{array}
  \right),
\end{eqnarray}
\begin{eqnarray}
  B_{\mu\nu}(x)&=&
  \left(
    \begin{array}{cccc}
      \frac{R^2+r^2}{R^2\sigma^2} & -\frac{r ct}{R^2\sigma^2} & 0 & 0\\
      -\frac{r ct}{R^2\sigma^2} & -\frac{R^2- c^2t^2}{R^2\sigma^2} & 0 & 0\\
      0 & 0 & -\frac{r^2}{\sigma} & 0\\
      0 & 0 & 0 & -\frac{r^2\sin^2\theta}{\sigma}
    \end{array}
  \right),
\end{eqnarray}
where $\sigma=\frac{R^2-c^2t^2+r^2}{R^2}$, and the following spherical coordinates expression of Beltrami metric has been used:
\begin{eqnarray}
 \nonumber ds^2_{Bel} &=& B_{\mu\nu}(x) dx^\mu dx^\nu \\
  &=&\frac{R^2(R^2+r^2)}{(R^2+r^2-c^2t^2)^2}c^2dt^2-\frac{2rR^2 ct}{(R^2+r^2-c^2t^2)^2}cdtdr
 \nonumber -\frac{R^2(R^2-c^2t^2)}{(R^2+r^2-c^2t^2)^2}dr^2\nonumber\\
  \label{Bel}&&-\frac{r^2R^2}{R^2+r^2-c^2t^2}(d\theta^2+\sin^2\theta d\phi^2) .\label{bspherical}
\end{eqnarray}
where subindex {\it Bel} means Beltrami.
Under reference system transformation between $\{x_N^\mu\}$ and $\{x^\mu\}$ the transformation from $g^{(0)}_{\alpha\beta}(x_N)$ to $B_{\mu\nu}(x)$ reads
\begin{eqnarray}\label{eq3-2-3}
g^{(0)}_{\alpha\beta}(x_N)\rightarrow B_{\mu\nu}(x)=\frac{\partial x_N^\alpha}{\partial x^\mu}\frac{\partial x_N^\beta}{\partial x^\nu}g^{(0)}_{\alpha\beta}(x_N),
\end{eqnarray}
which can be rewritten in matrix form,
\begin{eqnarray}\label{eq:trans}
  \mathcal{B}=T'g^{(0)}T,
\end{eqnarray}
where matrices $\mathcal{B}\equiv \{B_{\mu\nu}(x)\},\;g^{(0)}\equiv\{g^{(0)}_{\alpha\beta}(x_N)\},\;T\equiv \{\frac{\partial x_N^\beta}{\partial x^\nu}\}$, and $T'$ is the transpose of the matrix $T$.

To simplify the problem, we assume $T$ has the form of
\begin{eqnarray}
  T&=&
  \left(
    \begin{array}{cccc}
      \frac{\partial t_N}{\partial t}
      & \frac{\partial t_N}{\partial r} & 0 & 0\\
      \frac{\partial r_N}{\partial t}
      & \frac{\partial r_N}{\partial r} & 0 & 0\\
      0 & 0 & \frac{\partial \theta_N}{\partial \theta} & 0\\
      0 & 0 & 0 & \frac{\partial \phi_N}{\partial \phi}
    \end{array}
  \right).
\end{eqnarray}

Then from (\ref{eq:trans}) we have
\begin{eqnarray}
  \label{eq:x}
  \left\{
    \begin{array}{rll}
      (1-\frac{r_N^2}{R^2}) (\frac{\partial t_N}{\partial t})^2
      -\frac{(\frac{\partial r_N}{\partial t})^2}{1-\frac{r_N^2}{R^2}}
      &=\frac{R^2+r^2}{R^2\sigma^2}\\
      (1-\frac{r_N^2}{R^2}) \frac{\partial t_N}{\partial t} \frac{\partial t_N}{\partial r}
      -\frac{\frac{\partial r_N}{\partial t} \frac{\partial r_N}{\partial r}}{1-\frac{r_N^2}{R^2}}
      &=-\frac{r ct}{R^2\sigma^2}\\
      (1-\frac{r_N^2}{R^2}) (\frac{\partial t_N}{\partial r})^2 -\frac{(\frac{\partial r_N}{\partial r})^2}{1-\frac{r_N^2}{R^2}}
      &=-\frac{R^2- c^2t^2}{R^2\sigma^2}\\
      r_N^2 (\frac{\partial \theta_N}{\partial \theta})^2 &=\frac{r^2}{\sigma}
      &\Rightarrow (\frac{\partial \theta_N}{\partial \theta})^2=\frac{r^2}{\sigma r_N^2}\\
      r_N^2\sin^2\theta_N (\frac{\partial \phi_N}{\partial \phi})^2 &=\frac{r^2\sin^2\theta}{\sigma}
      &\Rightarrow (\frac{\partial \phi_N}{\partial \phi})^2=\frac{r^2\sin^2\theta}{\sigma r_N^2\sin^2 \theta_N}.
    \end{array}
  \right.
\end{eqnarray}

Assume $\frac{\partial t_N}{\partial r}=0$,
to solve $\frac{\partial t_N}{\partial t}$, $\frac{\partial r_N}{\partial t}$
and $\frac{\partial r_N}{\partial r}$, let
\begin{equation}
  A=\frac{R^2+r^2}{R^2\sigma^2},\quad
  B=-\frac{R^2- c^2t^2}{R^2\sigma^2},\quad
  C=-\frac{r ct}{R^2\sigma^2},\quad
  F=1-\frac{r_N^2}{R^2},
\end{equation}
the first three equations of (\ref{eq:x}) can be reduced to
\begin{eqnarray}
F (\frac{\partial t_N}{\partial t})^2 - \frac{(\frac{\partial r_N}{\partial t})^2}{F} &=&A,\\
-\frac{\frac{\partial r_N}{\partial t} \frac{\partial r_N}{\partial r}}{F} &=&C,\\
-\frac{(\frac{\partial r_N}{\partial r})^2}{F} &=&B.
\end{eqnarray}

The solution is
\begin{eqnarray}
\label{x11}  (\frac{\partial r_N}{\partial r})^2&=&-B F =
  \frac{1}{\sigma^2}\left(1-\frac{c^2t^2}{R^2}\right) \left(1-\frac{r_N^2}{R^2}\right),\\
\label{x10}  \frac{\partial r_N}{\partial t}&=&\frac{-C F}{\sqrt{-B F}}
  =\frac{r ct}{R^2 \sigma^{3/2}} \sqrt{\frac{1-\frac{r_N^2}{R^2}}{1-\frac{c^2t^2}{R^2}}},\\
\label{x00}  (\frac{\partial t_N}{\partial t})^2&=&\frac{A}{F}-\frac{C^2}{B F}
  =\frac{1}{\sigma \left(1-\frac{c^2t^2}{R^2}\right) \left(1-\frac{r_N^2}{R^2}\right)}.
\end{eqnarray}

From (\ref{x11}) (note $1-\frac{c^2t^2}{R^2}=\sigma -\frac{r^2}{R^2}$),
and $r_N' \equiv \frac{\partial r_N}{\partial r}$,
\begin{equation}
(\sqrt{\sigma} r_N')^2=(1- \frac{r^2}{R^2 \sigma})(1-\frac{r_N^2}{R^2}),
\end{equation}
we get a special solution of the coordinate transformation between non-inertial and inertial systems
\begin{equation}
r_N=\frac{r}{\sqrt{\sigma}}.
\end{equation}

From (\ref{x00}),
\begin{equation}
\frac{\partial t_N}{\partial t}
  =\sqrt{\frac{1}{\left(1-\frac{c^2t^2}{R^2}\right)
  \left(\sigma-\frac{(\sqrt{\sigma} r_N)^2}{R^2}\right)}}
  =\frac{1}{1-\frac{c^2t^2}{R^2}}.
\end{equation}

Finally, we get a coordinate transformation to inertial spacetime frame
\begin{eqnarray}
\label{45}  r_N&=&\frac{r}{\sqrt{1+\frac{r^2-c^2t^2}{R^2}}},\\
\label{46}  t_N&=&\int\frac{dt}{1-\frac{c^2t^2}{R^2}} = \frac{R}{c} \arctan{\frac{ct}{R}},\\
\label{47}  \theta_N&=&\theta,\\
\label{48}  \phi_N&=&\phi.
\end{eqnarray}
It's consistant with \cite{Lu74}.

\subsection{Schwarzschild-Beltrami-de Sitter Metric}\label{subsec:33}
Under the transformation $T$ of equations (\ref{45})$-$(\ref{48}),
the S-dS metric (\ref{3-1}) transforms to S-BdS metric:
\begin{eqnarray}\label{49}
g^{(M)}_{\mu\nu}(x_N)\rightarrow {}^\mathfrak{B}g^{(M)}_{\mu\nu}(x)=\frac{\partial x_N^\alpha}{\partial x^\mu}\frac{\partial x_N^\beta}{\partial x^\nu}g^{(M)}_{\alpha\beta}(x_N).
\end{eqnarray}
Substituting (\ref{45})$-$(\ref{48}) into (\ref{49}), we finally obtain the desired
S-BdS metric as follows
\begin{eqnarray}\nonumber
ds^2&=&{}^\mathfrak{B}g_{\mu\nu}^{(M)}(x)dx^\mu dx^\nu \\
&=&
\left(\frac{1-\frac{r^2}{R^2 \sigma }-\frac{2 G M \sqrt{\sigma}}{r}}
{\left(1-\frac{c^2t^2}{R^2}\right)^2}
-\frac{\frac{r^2 c^2t^2}{R^4}}
{\left(1-\frac{r^2}{R^2 \sigma }-\frac{2 G M \sqrt{\sigma }}{r}\right) \sigma ^3}\right)
c^2dt^2 \nonumber\\
&& -2 \frac{\frac{r ct}{R^2}\left(1-\frac{c^2t^2}{R^2}\right)}
{\left(1-\frac{r^2}{R^2 \sigma }-\frac{2 G M \sqrt{\sigma }}{r}\right) \sigma ^3}
cdt dr \nonumber\\
\label{bdss}&& -\frac{\left(1-\frac{c^2t^2}{R^2}\right)^2}
{\left(1-\frac{r^2}{R^2 \sigma }-\frac{2 G M \sqrt{\sigma }}{r}\right) \sigma ^3}
dr^2
-\frac{r^2}{\sigma}(d\theta^2+\sin^2\theta d\phi^2).
\end{eqnarray}
This is a new metric of dS-GR, and serves as main result of this paper.
It is a metric written in inertial Beltrami coordinates.
It is straightforward to check that it satisfies the Einstein field equation of dS-GR in empty spacetime, (\ref{13}). It is also easy to see that when $R\rightarrow \infty$, ${}^\mathfrak{B} g^{(M)}_{\mu\nu}(x)$ of (\ref{bdss}) coincides with Schwarzschild metric of (\ref{2});
 when $M\rightarrow 0$, it coincides with Beltrami metric of (\ref{bspherical});
 and when $R\rightarrow \infty$ and $M\rightarrow 0$,
it goes back to Minkowski metric of (\ref{3}).

\section{Summary and Discussion}\label{sec:con}

In this paper we start with a brief review to the de Sitter invariant special relativity (dS-SR), and construct de Sitter general relativity (dS-GR) via localizing the global de Sitter spacetime symmetry, which is equivalent to the GR with a cosmology constant $\Lambda=3/R^2$. We emphasized that the Beltrami metric $B_{\mu\nu}$
in corresponding Beltrami coordinates
plays an essential role to characterize the dS-spacetime with inertial reference frames in both dS-SR and dS-GR. Namely the motions of free particles in
Beltrami coordinates are inertial (i.e., along straight or say geodesic line with uniform velocity). Existence of inertial reference systems is the foundation of special relativity. And existence of local inertial reference systems is one of GR-principles. Physically, it is useful and meaningful to find out GR-solutions for empty spacetime, which approach to the metric of the inertial system when the gravity vanishes.

After reexamining the Schwarzschild-de Sitter (S-dS) metric $g_{\mu\nu}^{(M)}$ existed in literatures sine 1918, we find that when the gravity arisen from ``solar mass'' $M$ from $g_{\mu\nu}^{(M)}$ disappears, the metric $g_{\mu\nu}^{(0)}$ is not equal to the Beltrami metric $B_{\mu\nu}$, and then the motion of a free particle in $g_{\mu\nu}^{(0)}-$spacetime is non-inertial, or violates the so called inertial motion law. This means that the existed S-dS metric $g_{\mu\nu}^{(M)}$ is in non-inertial $g_{\mu\nu}^{(0)}-$spacetime. So $g_{\mu\nu}^{(M)}$ describes some mixing effects of gravity and inertial-force, instead of a purely gravity effect arisen from ``solar mass'' $M$. As is well known that,
in appropriate local inertial Minkowski spacetime coordinates
in which the inertial motion law holds,
the ordinary Schwarzschild metric in usual GR is a description of pure gravities due to $M$.
Generally, the predictions in inertial reference systems play essential role to clarify the corresponding physics conceptually. Almost all valuable predictions for experiments in particle physics, for instance, are formulated in expressions in inertial systems. In GR, the remarkable calculations of the motion of a planet in the gravitational field of the Sun, and of the bent of the light ray under the influence of the gravity of the Sun, and etc are achieved in terms of Schwarzschild metric which is just a GR-prediction in inertial framework.
Therefore, it is necessary and useful to find out the Schwarzschild-de Sitter metric
written in inertial Beltrami coordinates in dS-GR, or in short
Schwarzschild-Beltrami-de Sitter (S-BdS) metric ${}^\mathfrak{B}g_{\mu\nu}^{(M)}(x)$.
We provide the result as in (\ref{bdss}). This is the main result of this paper. We would like to mention that if parameter $R$ is finite (instead of infinite), both the motion of a planet in the gravitational field of the Sun and the bent of the light ray under the influence of the Sun will have to use ${}^\mathfrak{B}g_{\mu\nu}^{(M)}(x)$ to do such calculations. This project is in progress \cite{Deng}.
In addition, the dS-black hole physics should also be based on ${}^\mathfrak{B}g_{\mu\nu}^{(M)}(x)$ instead of the former $g_{\mu\nu}^{(M)}(x_N)$. However, it goes beyond the scope of this paper, and  remains to be open at present stage.

Finally, we would like to mention that ${}^\mathfrak{B}g_{\mu\nu}^{(M)}(x)$ is a time dependent metric. Since such a dependence is via a ratio of $c^2t^2/R^2$, and $R$ is a cosmologic huge length scale \cite{peebles}, usually $c^2t^2/R^2<<1$ and ordinary Schwarzschild metric in GR could be thought as a leading approximation of ${}^\mathfrak{B}g_{\mu\nu}^{(M)}(x)$.

\begin{center} {\bf ACKNOWLEDGMENTS}
\end{center}
{This work is partially supported by National Natural Science Foundation of China
under Grant No.~10975128 and No.~11031005, by Chinese Universities Scientific Fund
under Grant No.~WK0010000030, and by the Wu Wen-Tsun Key Laboratory of Mathematics
at USTC of Chinese Academy of Sciences.}

\appendix
\section{Beltrami Metric and de Sitter Invariant Special Relativity}
In this Appendix we present some explicit calculations related to de Sitter invariant Special Relativity (dS-SR), and some interpretations to reference \cite{Ours}.

\begin{enumerate}
\item Beltrami metric:

We derive the expression of Beltrami metric (\ref{5}) in the text.
We consider a 4-dimensional pseudo-sphere (or hyperboloid) $\mathcal{S}_\Lambda$ embedded in a 5-dimensional Minkowski spacetime with metric $\eta_{AB} =diag(1,-1,-1,-1,-1)$:
\bea\nn
\mathcal{S}_\Lambda :&&\eta_{AB}\xi^A\xi^B=-R^2,\\
\la{3-1app}&& ds^2=\eta_{AB}d\xi^Ad\xi^B,
\eea
where index $A,\;B=\{0,1,2,3,5\} $, $R^2:=3\Lambda^{-1}$ and $\Lambda$ is the cosmological constant.
$\mathcal{S}_\Lambda$ is also called de Sitter pseudo-spherical surface with radii $R$. Defining
\bea\la{3-2}
x^\mu:=R{\xi^\mu\over \xi^5},~~{\rm with}~~\xi^5\neq 0,~{\rm and}~\mu=\{0,1,2,3\}.
\eea
and treating $x^\mu$ are Cartesian-type coordinates of a 4-dimensional spacetime with metric $g_{\mu\nu}(x)\equiv B_{\mu\nu}(x)$, denoting this 4-dimensional spacetime as $\mathcal{B}_\Lambda$ (call it Beltrami spacetime), we derive $B_{\mu\nu}(x)$ by means of the geodesic projection of $\{\mathcal{S}_\Lambda\mapsto \mathcal{B}_\Lambda\}$ (see Figure 1).
\begin{figure}[ht]
\begin{center}%\vskip-0.5in
\includegraphics[width=0.5\textwidth]{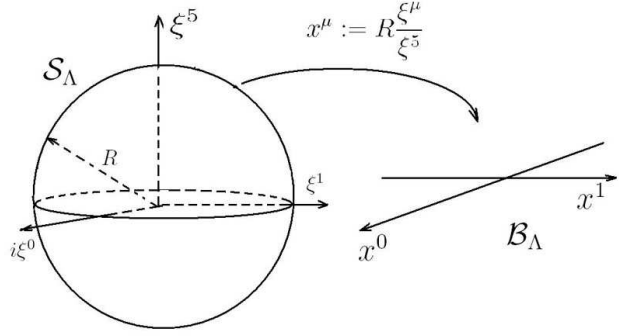}
%\includegraphics[scale=0.4]{dS4}
%\vskip-3in
\caption{\label{Fig1} \small Sketch of the geodesic projection from de Sitter pseudo-spherical surface $\mathcal{S}_\Lambda$ to the Beltrami spacetime $\mathcal{B}_\Lambda$ via Eq.(\ref{3-2}).}
\end{center}
\end{figure}
From the definition (\ref{3-1app}), we have
\bea\nn
ds^2&=&\eta_{AB}d\xi^A d\xi^B|_{\xi^{A,B}\in \mathcal{S}_\Lambda}\\
\nn&=&\eta_{\mu\nu}d\xi^\mu d\xi^\nu-(d\xi^5)^2\\
\la{3-3}&:=&B_{\mu\nu}(x)dx^\mu dx^\nu .
\eea
Since $\xi^{A,B}\in \mathcal{S}_\Lambda $, and from (\ref{3-2}) and (\ref{3-1app}), it is easy to obtain:
\bea\nn
&&\xi^\mu={x^\mu \over R}\xi^5,~~d\xi^\mu={1\over R}(\xi^5dx^\mu+x^\mu d\xi^5),~~ (\xi^5)^2={R^2\over \sigma(x)},\\
\nn && d\xi^5=\eta_{\mu\nu} {\xi^\mu\over \xi^5}d\xi^\nu={1\over R}\eta_{\mu\nu}x^\mu d\xi^\nu
={\eta_{\mu\nu}x^\mu dx^\nu\over \xi^5\sigma(x)^2},
\eea
where
\bea\la{333}
\sigma(x)= 1-{\eta_{\mu\nu}x^\mu x^\nu \over R^2}.
\eea
Substituting them into Eq.(\ref{3-3}), we have
\bea\nn
ds^2={\eta_{\mu\nu}dx^\mu dx^\nu\over \sigma(x)}+{(\eta_{\mu\nu}x^\mu dx^\nu)^2\over R^2 \sigma(x)^2}:= B_{\mu\nu}(x)dx^\mu dx^\nu.
\eea
Then, we obtain the Beltrami metric as follows
\bea\la{3-4}
B_{\mu\nu}(x)={\eta_{\mu\nu}\over \sigma(x)}+{\eta_{\mu\lambda}x^\lambda \eta_{\mu\rho}x^\rho\over R^2 \sigma(x)^2}.
\eea

\item Inertial reference coordinates and principle of relativity: \\
The first Newtonian law is the foundation of the relativity. This law claims that the free particle moves with uniform velocity and along straight line.
There exist systems of reference in which the first Newtonian motion law holds. Such reference systems are defined to be {\it inertial}. And the Newtonian motion law is always called {\it the inertial moving law}. If two reference systems move uniformly relative to each other, and if one of them is an inertial system, then clearly the other is also inertial.
Experiment, e.g., the observations in the Galileo-boat which moves uniformly, shows that the so-called {\it principle of relativity} is valid. According to this principle all the law of nature are identical in all inertial systems of reference.

\noindent {\it Theorem 1}: The motion of particle with mass $m_0$ and described by the following Lagrangian
\begin{equation}\label{A1}
L_{Newton}={1\over2}m_0\mathbf{v}^2={1\over2}m_0\dot{\mathbf{x}}^2
\end{equation}
satisfy the first Newtonian motion law, or the motion is {\it inertial}. In (\ref{A1}), the Cartesian expression
of the velocity is as follows
\begin{equation}\label{A2}
\mathbf{v}\equiv \dot{\mathbf{x}},~~{\rm and}~~\mathbf{x}=x^1\mathbf{i}+x^2\mathbf{j}+x^3\mathbf{k},
\end{equation}
where $\mathbf{i}\cdot\mathbf{i}=\mathbf{j}\cdot\mathbf{j}=\mathbf{k}\cdot\mathbf{k}=1$, and $\mathbf{i}\cdot\mathbf{j}=\mathbf{i}\cdot\mathbf{k}=\mathbf{j}\cdot\mathbf{k}=0$.

{\it Proof}: By means of the Euler-Lagrangian equation
\begin{equation}\label{A3}
{\pa L \over \pa x^i}={d \over dt} {\pa L \over \pa \dot{x}^i},~~{\rm or}~~{\pa L \over \pa \mathbf{x}}={d \over dt} {\pa L \over \pa \dot{\mathbf{x}}}
\end{equation}
(where $\pa /\pa \mathbf{x}\equiv \nabla:=(\pa /\pa x^1)\mathbf{i}+(\pa /\pa x^2)\mathbf{j}+(\pa /\pa x^3)\mathbf{k}$ and etc) and $L=L_{Newton}$ we obtain
\begin{equation}\label{A4}
\ddot{x}^i=0,~~~~\dot{x}^i=v^i=constant,~~{\rm or}~~\dot{\mathbf{x}}=\mathbf{v}=constant.~~~~QED.
\end{equation}

\noindent {\it Theorem 2}: The motion of particle in Minkowski spacetime described by
\begin{equation}\label{A5}
L_{Einstein}=-m_0c{ds\over dt}=-m_0c{\sqrt{\eta_{\mu\nu}dx^\mu dx^\nu}\over dt}=-m_0c^2\sqrt{1-{\dot{\mathbf{x}}^2\over c^2}}
\end{equation}
is inertial.

The proof is the same as above, because both $L_{Newton}$ and $L_{Einstein}$ are coordinates
$x^i$-independent. Generally, any $\mathbf{x}$-free and time $t$-free Lagrangian functions $L(\dot{\mathbf{x}})$ can always reach the result of (\ref{A4}). However,
when Lagrangian function is time-dependent that rule will become invalid.
A useful example is as follows:
\begin{eqnarray}\label{A6}
L_{\Lambda}(t,\mathbf{x},\dot{\mathbf{x}})
=-m_0c^2 \sqrt{3/\Lambda} \sqrt{3/\Lambda (c^2- \dot{\mathbf{x}}^2)-\mathbf{x}^2\dot{\mathbf{x}}^2+(\mathbf{x}\cdot\dot{\mathbf{x}})^2+c^2(\mathbf{x}-\dot{\mathbf{x}}t)^2  \over c^2 (3/\Lambda +\mathbf{x}^2-c^2t^2)^2},
\end{eqnarray}
where a constant $\Lambda\neq 0$. The stick-to-itive readers can verify the following identity via straightforward calculations from (\ref{A6}):
\bea\la{A7}
{\pa L_{\Lambda} \over \pa \mathbf{x}}={\pa \over \pa t} {\pa L_{\Lambda} \over \pa \dot{\mathbf{x}}}+\left(\dot{\mathbf{x}}\cdot{\pa \over \pa\mathbf{x}}\right) {\pa L_{\Lambda} \over \pa \dot{\mathbf{x}}}.
\eea
Noting that the Euler-Lagrange equation (\ref{A3}) reads
\bea\la{A8}
{\pa L_{\Lambda} \over \pa \mathbf{x}}={d \over dt} {\pa L_{\Lambda} \over \pa \dot{\mathbf{x}}} ={\pa \over \pa t} {\pa L_{\Lambda} \over \pa \dot{\mathbf{x}}}+\left(\dot{\mathbf{x}}\cdot{\pa \over \pa\mathbf{x}}\right) {\pa L_{\Lambda} \over \pa \dot{\mathbf{x}}}+ \left(\ddot{\mathbf{x}}\cdot{\pa \over \pa\dot{\mathbf{x}}}\right) {\pa L_{\Lambda} \over \pa \dot{\mathbf{x}}},
\eea
and substituting (\ref{A7}) to (\ref{A8}), we have
\bea\la{A9}
\left(\ddot{\mathbf{x}}\cdot{\pa \over \pa\dot{\mathbf{x}}}\right) {\pa L_{\Lambda} \over \pa \dot{\mathbf{x}}}=0.
\eea
Since
\bea\la{A10}
\|{\pa \over \pa\dot{\mathbf{x}}} {\pa L_{\Lambda} \over \pa \dot{\mathbf{x}}}\|\equiv
\det \left({\pa^2L_\Lambda\over \pa x^i\pa x^j}\right)\neq 0
\eea
we have
\begin{equation}\label{A11}
\ddot{\mathbf{x}}=0,~~~~\dot{\mathbf{x}}=\mathbf{v}=constant,
\end{equation}
which indicates that the particle motion described by Lagrangian function (\ref{A6}) is inertial, and the first Newton motion law holds. Thus, the corresponding inertial reference systems can be built. Noting
\bea\la{A12}
\lim_{\Lambda\rightarrow 0}L_{\Lambda}=L_{Einstein},
\eea
it is essential and remarkable that a new kind of Special Relativity based on $L_\Lambda$ (\ref{A6}) serving as an extension of the Einstein's Special Relativity (E-SR) may exist.

\item de Sitter invariant Special Relativity (dS-SR):\\
Following the Landau-Lifshitz formulation of Lagrangian \cite{Landau} (see (\ref{A5})), we examine the motion of free particle in the spacetime with Beltrami metric (\ref{3-4}). From Eq.(\ref{4}) in text
\begin{equation}\label{B1}
 L_{dS}=-m_0c \frac{ds}{dt}
 =-m_0c{\sqrt{B_{\mu\nu}(x)dx^\mu dx^\nu}\over dt}=-m_0c{\sqrt{B_{\mu\nu}(x)\dot{x}^\mu \dot{x}^\nu}},
 \end{equation}
we derive its expression in Cartesian coordinates.
Setting up the time
$t=x^0/c$,   $B_{\mu\nu}(x)$ can be rewritten as follows
\begin{eqnarray}\label{3-19app}
ds^2&=&B_{\mu\nu}(x) dx^\mu dx^\nu
=\widetilde{g}_{00}d(ct)^2+\widetilde{g}_{ij}\left[(dx^i+N^id(ct))
(dx^j+N^jd(ct))\right]\\ \nonumber &=& c^2 (dt)^2
\left[\widetilde{g}_{00} +\widetilde{g}_{ij}({1\over
c}\dot{x}^i+N^i) ({1\over c}\dot{x}^j+  N^j)\right],
\end{eqnarray}
where
\begin{eqnarray}\label{3-20}
\widetilde{g}_{00}&=&{R^2\over \sigma(x) (R^2-c^2t^2)},\\
\label{3-21} \widetilde{g}_{ij}&=&{\eta_{ij}\over \sigma (x)}+
{1\over
R^2\sigma(x)^2}\eta_{il}\eta_{jm}x^lx^m,\\
\label{3-22} N^i&=&{ctx^i \over R^2-c^2t^2}.
\end{eqnarray}
Substituting eqs.(\ref{3-19app})--(\ref{3-22}) into (\ref{B1}), we
obtain the Lagrangian for free particle in $ \mathcal{B}_{\Lambda} $:
\begin{equation}\label{3-23}
 L_{dS}=-m_0c^2 \sqrt{\widetilde{g}_{00} +\widetilde{g}_{ij}({1\over
c}\dot{x}^i+N^i) ({1\over c}\dot{x}^j+  N^j)}.
 \end{equation}
By using Cartesian notations (\ref{A2}) and expressions of (\ref{333}) (\ref{3-20}) (\ref{3-21}) (\ref{3-22}), the explicit expression of Lagrangian (\ref{3-23}) is:
\bea\nn
L_{dS}&\hskip-0.1in=&\hskip-0.1in -m_0c^2\left[{R^4\over (R^2+\mathbf{x}^2-c^2t^2)(R^2-c^2t^2)}\right.\\
\nn && +\left.{-R^2\over R^2+\mathbf{x}^2-c^2t^2}\right.\left({\dot{\mathbf{x}}^2\over c^2}+{c^2t^2\mathbf{x}^2\over (R^2-c^2t^2)^2}+{2t(\mathbf{x}\cdot\dot{\mathbf{x}})\over R^2-c^2t^2}\right)\\
\nn && +\left.{R^2\over (R^2+\mathbf{x}^2-c^2t^2)^2}\left({\dot{\mathbf{x}}\cdot\mathbf{x}\over c}+{ct\mathbf{x}^2\over R^2-c^2t^2}\right)^2\right]^{1/2}\\
\la{3-26} &=& -m_0c^2 R \sqrt{R^2 (c^2- \dot{\mathbf{x}}^2)-\mathbf{x}^2\dot{\mathbf{x}}^2+(\mathbf{x}\cdot\dot{\mathbf{x}})^2+c^2(\mathbf{x}-\dot{\mathbf{x}}t)^2  \over c^2 (R^2+\mathbf{x}^2-c^2t^2)^2},
\eea
where $\mathbf{x}^2=(\mathbf{x}\cdot\mathbf{x})$. Noting Eq.(\ref{12}) in text, $R^2=3/\Lambda$, and comparing $L_{dS}$ with $L_{\Lambda}(t,\mathbf{x},\dot{\mathbf{x}})$ of (\ref{A6}), we find
\bea\la{B10}
L_{dS}=L_{\Lambda}(t,\mathbf{x},\dot{\mathbf{x}})=-m_0c{\sqrt{B_{\mu\nu}(x)\dot{x}^\mu \dot{x}^\nu}},
\eea
which is the Lagrangian for free particle mechanics of dS-SR. Since (\ref{A12}), when $|R|\rightarrow \infty$, the dS-SR goes back to E-SR.

\item de Sitter transformation to preserve Beltrami metric $B_{\mu\nu}$:\\
In the text, Eq.(\ref{8}) represents the de Sitter transformation to preserve Beltrami metric $B_{\mu\nu}$. When  space rotations were neglected temporarily  for simplify, the
transformation both due to a Lorentz-like boost and a
space-transition in the $x^1$ direction with  parameters
$\beta=\dot{x}^1/c$ and $a^1$ respectively and due to a time
transition with  parameter $a^0$ can be explicitly written as
follows:
\begin{eqnarray}\label{general transformation}
\begin{array}{rcl}
t\rightarrow \tilde{t}&=& \frac{\sqrt{\sigma(a)}}{c \sigma(a,x)}
\gamma \left[ct-\beta x^1-a^0+ \beta a^1 +\frac{a^0-\beta
a^1}{R^2}\frac{a^0 ct-a^1 x^1-(a^0)^2 +(a^1)^2 }
{ \sigma(a)+\sqrt{\sigma(a)}} \right] \\
 x^1\rightarrow \tilde{x}^1&=& \frac{\sqrt{\sigma(a)}}{
\sigma(a,x)}\gamma \left[ x^1-\beta ct +\beta a^0 -a^1 +\frac{a^1-
\beta a^0}{R^2}
\frac{a^0 ct-a^1 x^1-(a^0)^2 +(a^1)^2}{ \sigma(a)+\sqrt{\sigma(a)}}\right]\\
 x^2\rightarrow
\tilde{x}^2&=&\frac{\sqrt{\sigma(a)}}{\sigma(a,x)}x^2 \\
 x^3\rightarrow
\tilde{x}^3&=&\frac{\sqrt{\sigma(a)}}{\sigma(a,x)}x^3
\end{array}
\end{eqnarray}
where $\gamma=1/\sqrt{1-\beta^2}$. It is easy to check when $R\rightarrow \infty$ the above
transformation goes back to Poincar\'e transformation (or
inhomogeneous Lorentz group $ISO(1,3)$ transformation) in E-SR.

\item Conserved Noether charges of $SO(4,1)$ of dS-SR:\\
The external spacetime symmetry of dS-SR is $SO(4,1)$. According to Neother theorem, the corresponding 10-Noether charges are energy $E$, momentums $p^i$, boost charges $K^i$ and angular-momentums $L^i$. All have been derived in \cite{Ours}. The results are as follows
\begin{eqnarray}\label{503a}
\begin{array}{rcl}
  &&{\rm{ Noether}\;charges\;for\;Lorentz\;boost:\;} ~~
 K^i=m_0 \Gamma c (x^i- t \dot{x}^i) \\
 &&{\rm
Charges\;for\;space-transitions\;(momenta):}~~~  p^i=m_0 \Gamma \dot{x}^i, \\
 &&{\rm Charge\;for\;time-transition\;(energy): }~~~
 E= m_0 c^2 \Gamma \\
&&{\rm Charges\;for\;rotations\;in\;space\;(angular momenta):}~~~
L^i = \epsilon^{i}_{jk}x^{j}p^{k},
\end{array}
\end{eqnarray}
where the Lorentz factor of dS-SR is:
\begin{eqnarray} \label{new parameter}
 \Gamma ={1\over \sqrt{1-{\dot{\mathbf{x}}^2\over c^2}+{(\mathbf{x}\cdot \dot{\mathbf{x}})^2 -\mathbf{x}^2\dot{\mathbf{x}}^2\over c^2R^2}+{(\mathbf{x}-\dot{\mathbf{x}}t)^2\over R^2}}}.
\end{eqnarray}
It can be checked that $\dot{E}=\dot{p^i}=\dot{K^i}=\dot{L^i}=0$ under the equation of motion $\ddot{x}^i=0$ (or $\ddot{\mathbf{x}}=0$). \cite{Ours}

\end{enumerate}


\begin{thebibliography}{99}
\bibitem{peebles} P.J.E. Peebles, B. Ratra, Rev. Mod. Phys. {\bf 75} (2003) 559;
T. Padmanabhan, Phys. Rep. {\bf 380} (2003) 235.

\bibitem{look} K.H. Look (Q.K. Lu), {\it Why the Minkowski metric must be used?}, (1970), unpublished.
\bibitem{Lu74} K.H. Look, C.L. Tsou (Z.L. Zou) and H.Y. Kuo (H.Y. Guo), {\it Acta Physica Sinica}, {\bf 23} (1974) 225 (in Chinese).

\bibitem{Guo1} H.Y. Guo, C.G. Huang, Z. Xu, and B. Zhou, Phys. Lett. {\bf A 331} (2004) 1;
Mod. Phys. Lett. {\bf A 19} (2004) 1701; Chin. Phys. Lett. {\bf 22} (2005) 2477; arXiv:hep-th/0405137;
H.Y. Guo, C.G. Huang and B. Zhou, arXiv:hep-th/0404010.
\bibitem{Guo2} Y. Tian, H.Y. Guo, C.G. Huang, Z. Xu and B. Zhou, Phys. Rev. {\bf D 71} (2005) 044030.
\bibitem{Guo3} H.Y. Guo, Class. Quant. Grav. {\bf 24} (2007) 4009, arXiv:gr-qc/0703078.
\bibitem{Guo4} H.Y. Guo, C.G. Huang, H.T. Wu, Phys. Lett. {\bf B 663} (2008) 270.

\bibitem{Ours} M.L. Yan, N.C. Xiao, W. Huang, S. Li,
%{\it Hamiltonian Formalism of the de-Sitter Invariant Special Relativity},
Commun. Theor. Phys. {\bf 48} (2007) 27, arXiv:hep-th/0512319.
\bibitem{Yan1} S.X. Chen, N.C. Xiao, Mu-Lin Yan,
%{\it Variation of the Fine-Structure Constant from the de Sitter Invariant Special Relativity},
Chinese Phys. {\bf C 32} (2008) 612, arXiv:astro-ph/0703110.
\bibitem{Yan2} M.L. Yan,
%{\it Hydrogen Atom and Time Variation of Fine-Structure Constant},
Commun. Theor. Phys. {\bf 57} (06) 930-952 (2012), arXiv:1004.3023.
\bibitem{Yan3} M.L. Yan, S. Hu, W. Huang and N.C. Xiao,
%{\it On determination of the geometric cosmological constant and OPERA experiment of superluminal neutrinos},
Mod. Phys. Lett. {\bf A27} (2012) 1250041.

\bibitem{Landau} L.D. Landau and E.M. Lifshitz, {\it The Classical Theory of Fields}, (Translated from Russian by M. Hamermesh), Pergamon Press, Oxford (1987).
\bibitem{Utiyama} R. Utiyama, Phys. Rev. {\bf 101} (1956) 1597.
\bibitem{Kibble} T.W.B. Kibble, J. Math. Phys. {\bf 2} (1961) 212.

\bibitem{Kottler} F. Kottler,
%\"{U}ber die physikalischen Grundlagen der Einsteinschen Gravitationstheorie,
Ann. Physik {\bf 56} (361), 401-462 (1918);
H. Weyl,
%\"{U}ber die statischen, kugelsymmetrischen L\"{o}sungen von Einsteins kosmologischen Gravitationsgleichungen,
Phys. Z. {\bf 20}, 31-34 (1919);
E. Trefftz,
%Das statische Gravitationsfeld zweier Massenpunkte in der Einsteinschen Theorie,
Mathem. Ann. {\bf 86}, 317-326 (1922).
%\bibitem{Nariai} H. Nariai, {\it ``On some static solutions of Einstein¡¯s gravitational field equations in a spherically symmetric case"}, Sci. Rep. Tohoku Univ. {\bf 34} 160 (1950);
% {\it ``On a new cosmological solution of Einstein¡¯s field equations of gravitation"}, Sci. Rep. Tohoku Univ. {\bf 35}, 62 (1951).
\bibitem{Deng} Y. Deng, et al., (in progress).

\end{thebibliography}
\end{document}